\documentclass[prb,aps,preprint,epsfig]{revtex4}
\usepackage{graphicx}
\usepackage{dcolumn}
\usepackage{bm}

\begin{document}

\title {Localized reversible nanoscale phase separation in Pr$_{0.63}$Ca$_{0.37}$MnO$_3$ single
crystal using a scanning tunneling microscope tip}

\author{Sohini
Kar\footnote[1]{electronic mail:sohini@physics.iisc.ernet.in} }
\address{Department of Physics, Indian Institute of Science, Bangalore 560 012, India.}
\author{A. K. Raychaudhuri\footnote[2]{On lien from Department of Physics, Indian Institute
of Science, Bangalore 560 012, India; electronic mail:
arup@bose.res.in}}
\address{S.N.Bose National Centre for Basic Sciences, Salt Lake, Kolkata 700 098, India}


\begin{abstract}
We report the destabilization of the charge ordered insulating
(COI) state in a localized region of Pr$_{0.63}$Ca$_{0.37}$MnO$_3$
single crystal by current injection using a scanning tunneling
microscope tip. This leads to controlled phase separation and
formation of localized metallic nanoislands in the COI matrix
which have been detected by local tunneling conductance mapping.
The metallic regions thus created persist even after reducing the
injected current to lower values. The original conductance state
can be restored by injecting a current of similar magnitude but of
opposite polarity. We thus achieve reversible nanoscale phase
separation that gives rise to the possibility to ``write, read and
erase" nanosized conducting regions in an insulating matrix with
high spatial resolution.

\end{abstract}


\maketitle

     Rare-earth manganites having the perovskite structure
R$_{1-x}$A$_x$MnO$_3$, where \textit{R} is a rare earth ion and
\textit{A} is a divalent alkaline-earth ion, have been studied in
great detail \cite{cnr, dagotto} for the rich variety of distinct
phases they exhibit. Of particular interest is the charge ordered
insulating (COI) phase which is found to be unstable under various
external perturbations such as magnetic field \cite{kuwahara},
electric field and current\cite{ayan, asamitsu} (effects of which
are also observed in manganites which are not charge ordered
\cite{himanshu,wu,gao,ahn}) and also by size reduction to few tens
of nanometers \cite{tapati}. Normally, the phase that is created
by destabilization of the COI phase is the ferromagnetic metallic
(FMM) ground state. Thus these materials are attractive candidates
for various switching and sensing devices. The creation of FMM
filaments in a bulk single crystal of a COI manganite by using a
high current has been shown before\cite{ayan}. In the present
experiment, we demonstrate localized destabilization of the COI
phase by current injection using a scanning tunneling microscope
(STM) tip over a small area with size $\approx 25-30$nm. This
phenomenon gives rise to the possibility of patterning (writing)
nanosized metallic regions by controlled phase separation on an
otherwise insulating background. This process can be reversed
(erased) and the metallic region can be switched back to the COI
state at will by passing a current of reverse polarity and
comparable magnitude to the writing current. Hence, we are able to
achieve reversible and controlled phase separation over nanoscopic
length scales without altering the topography or structure of the
surface.

     The sample used in our experiments was a single crystal of
Pr$_{0.63}$Ca$_{0.37}$MnO$_3$ (PCMO) grown by float zone
technique. The sample has been well characterized and shows a CO
transition at $T_{CO}$= 235K\cite{ayan}. The variable temperature
ultra-high vacuum STM used was constructed in our laboratory and
we used an SPM100 controller (RHK Technology Inc., USA). The STM
was pumped to a base pressure of better than 1$\times10^{-8}$ Torr
after which cryopumping takes over. The STM tips were mechanically
formed Pt/Rh (87:13) wires. The crystal was freshly broken to
create a clean surface for STM measurements.

     In the COI phase, there exists a gap in the charge excitation spectra
($\triangle$). The gap collapses when the COI state is
destabilized into a metallic state as has been shown in magnetic
field induced destabilization \cite{amlan1}. We first measured
$\triangle$ through scanning tunneling spectroscopy (STS). The
tunnel current was stabilized at 1nA at a bias of 0.7V. The bias
was then swept between $\pm{0.75}$V and the tunnel current-voltage
($I-V$) spectra were recorded. The $dI/dV-V$ curves were evaluated
from the tunneling spectra and $\triangle$ was evaluated taking
into consideration the finite temperature effect. In figure 1, we
show $\triangle$ as a function of temperature. The CO temperature
can be clearly identified from the temperature dependence of the
gap. Above $T_{CO}$ the gap is $\approx 0.05eV$, as is generally
observed in the polaronic insulating states of other manganites
\cite{sohini ssc}. We then measured $\triangle$ for higher tunnel
currents in a step by step process by bringing the tip closer to
the sample. Higher tunnel currents were stabilized at the same
point for the same bias and the $I-V$ spectra were recorded again.
This process was repeated for various tunnel currents and the
$dI/dV-V$ curves were evaluated from the tunneling spectra. It was
found that the gap decreases linearly with increasing tunnel
current. The higher tunnel currents thus reduce the CO gap leading
to destabilization of the CO state, as can be seen from the inset
of figure 1 where we show the $dI/dV-V$ curves taken at 152K for
tunnel currents of 1nA and 4nA. The destabilization typically
occurs around a tunnel current of 4-5nA. This information is
important and is used in deciding the ``write-read-erase" cycle as
will be shown below.

\begin{figure}
\begin{center}
\includegraphics[width=6.5cm]{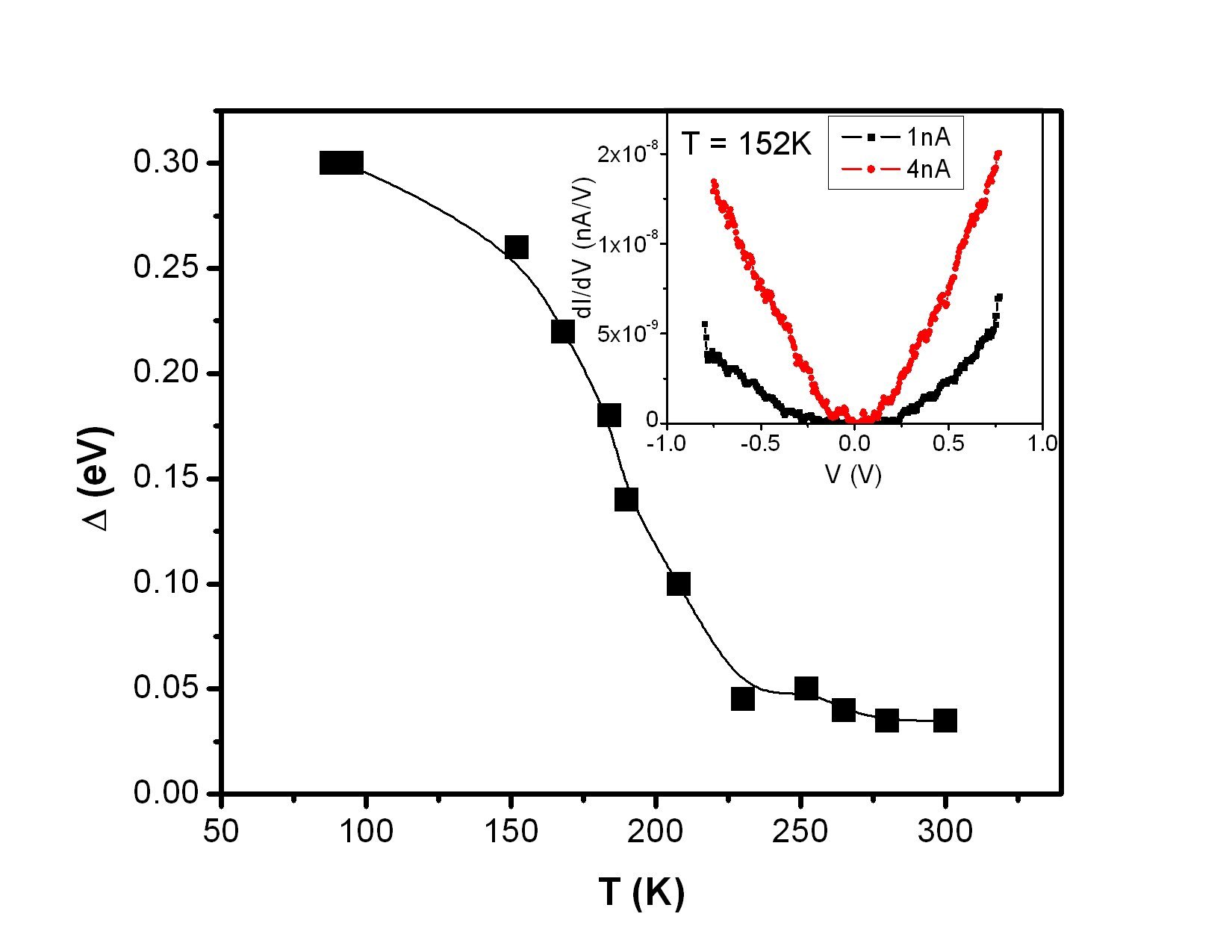}
\end{center}
\vspace{-0.7cm} \caption{The CO gap ($\triangle$), as a function
of temperature. The inset shows an example of the tunneling
spectra ($dI/dV-V$) taken at 1nA and 4nA at 152K showing the
collapse of the CO gap at higher tunnel current}. \label{figure1}
\end{figure}

     The above observation enables us to use an STM tip to selectively destabilize a given CO region thus controlling
spatially the region that one would like to drive metallic in a
COI matrix. We refer to this as the ``writing" process. We
demonstrate this in fig 2 where we have carried out local
tunneling conductance mapping (LCMAP) over a given area at
$T=152K$ ($T<T_{CO}$). The LCMAP technique generates a
two-dimensional map of the tunneling conductance across the
surface for a fixed bias and at a fixed average tunnel current.
This is achieved by applying a small ac bias modulation over the
dc sample bias and using a lock-in amplifier to measure $dI/dV$
directly. One can simultaneously record topography as well as the
LCMAP of the same area. Fig 2(a) shows the LCMAP of a
100nm$\times$100nm area of the surface of PCMO, taken at a tunnel
current of 1nA. We then zoomed into an area marked by the arrow
and scanned an area of 25nm$\times$25nm using a tunnel current of
5nA. This was the ``writing" cycle.  We then zoomed out to the
original 100nm$\times$100nm scan area and recorded the LCMAP again
with a tunnel current of 1nA. This was the ``read" process that
established whether the chosen region has become more conducting.
As can be seen from fig2(b), there is a square area of about
30nm$\times$30nm which has significantly higher tunnel conductance
than the surrounding region. In our color scheme, the higher
tunneling conductance region appears brighter. The brighter region
has a conductance that is typically $\approx 3$ times higher than
the surrounding darker region. (The surrounding region is not
uniformly dark and has small bright pockets that already existed
in the pristine single crystal.)

     To ``erase" the conducting region created in the above process, we
injected a current of comparable magnitude but of opposite
polarity. We zoomed back into the 25nm$\times$25nm spot that was
written before and scanned the same area again using -5nA tunnel
current. When we recorded the zoomed out image at 1nA again, we
found that we had managed to erase the metallic region almost
completely thus reverting it back to a region of low tunneling
conductance, as can be seen in fig 2(c). This is a very important
observation as this phenomenon can be used to selectively write,
read and erase nanosized metallic domains (regions of high tunnel
conductance) in an insulating background (regions of lower
tunneling conductance) using an STM tip. The small region of high
conductance marked with an arrow in figure 2(a) was used as a
marker, so that when the erase process was done one could see that
the region had almost reduced to the original size. The
conductance of the starting marker patch was about 5.9nA/V(fig
2(a)) and after the erasing the conductance of the marker patch is
6.4nA/V. (fig 2(c)). We note that in this process the topography
had not changed.

\begin{figure}
\begin{center}
\includegraphics[width=8.5cm]{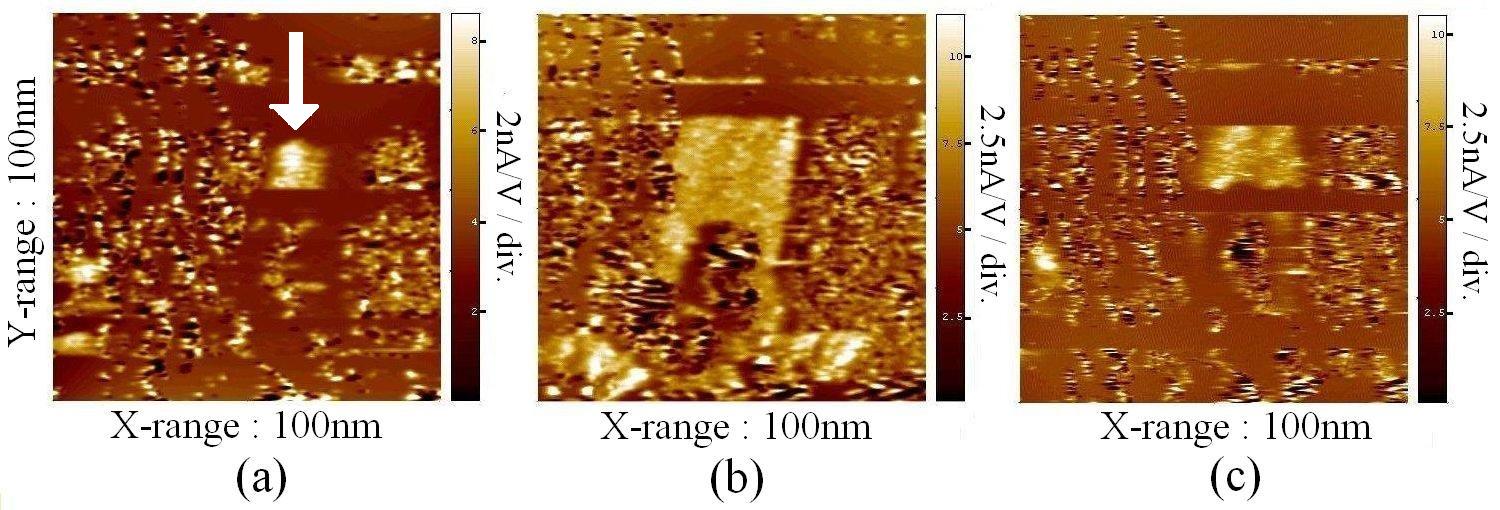}
\end{center}
\vspace{-0.7cm} \caption{Collection of LCMAPs taken at 152K
showing selective electronic phase transition in PCMO single
crystal. (a) LCMAP of 100nm$\times$100nm area taken at 1nA, (b)
LCMAP of the same area taken at 1nA tunnel current, after scanning
a 25nm$\times$25nm area near the center with tunnel current of 5nA
and (c) LCMAP of the same area taken at 1nA after scanning an area
of 25nm$\times$25nm at the center with -5nA tunnel current. (The
small thermal drift is not corrected.)} \label{figure1}
\end{figure}

We then carried out STS measurements to establish that the
nano-island created was a metallic region with no CO gap
($\triangle\rightarrow 0$). After the write process, we took
tunneling spectra at different points on the surface that included
the written region as well as the pristine regions. We found two
distinct $I-V$ curves as shown in figure 3. The inset shows the
corresponding $dI/dV-V$ curves. One set of $I-V$ curves taken on
the ``written" region shows metallic behavior with no gap
($\triangle\approx 0$) while the other set taken on the virgin
regions shows a finite gap ($\triangle\neq 0$) opening up in the
tunneling spectra. This shows that we have two distinct phases
coexisting, the COI phase with a finite gap and the metallic
region where a high tunnel current had been injected.

\begin{figure}
\begin{center}
\includegraphics[width=6.5cm]{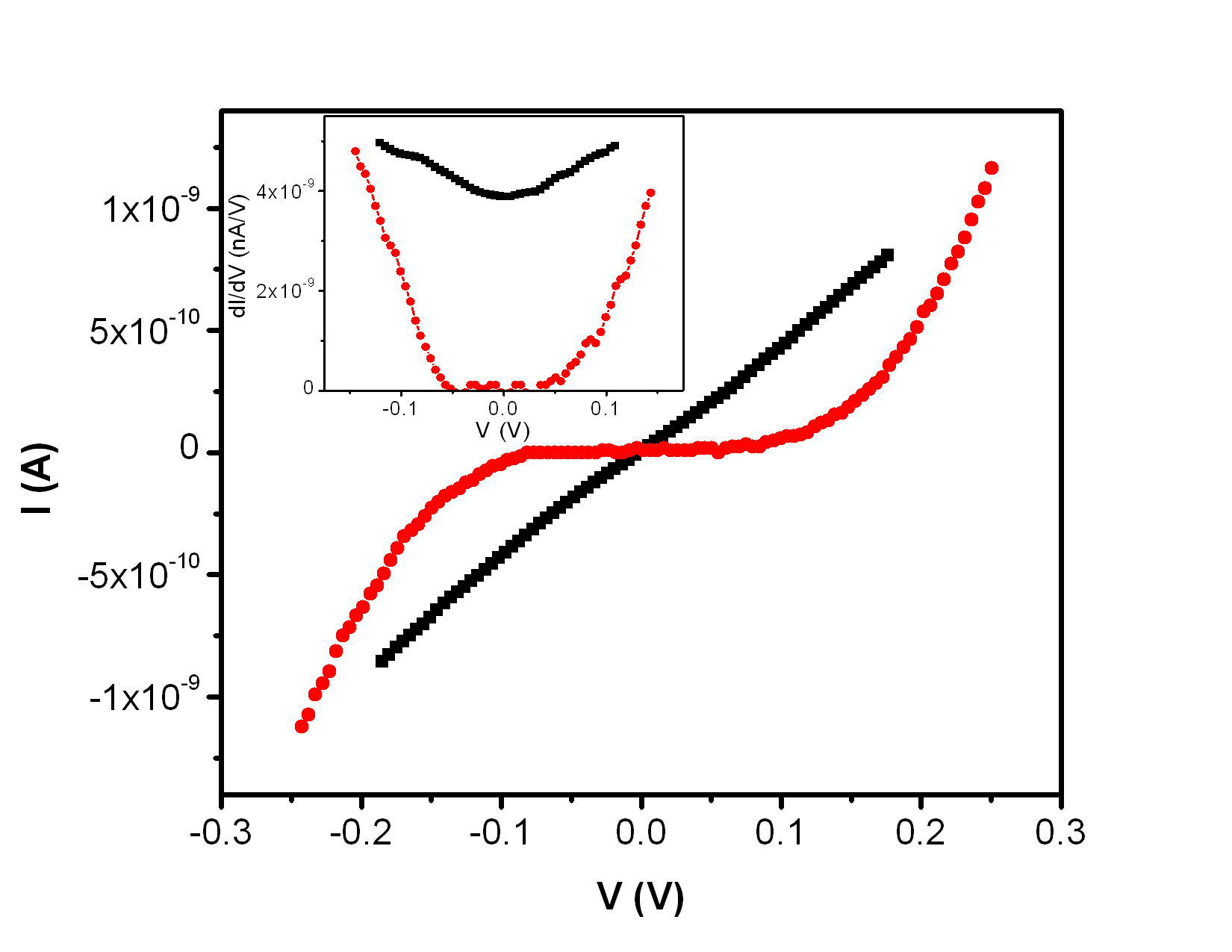}
\end{center}
\vspace{-0.7cm} \caption{Two distinct kinds of $I-V$s showing both
metallic (squares) and insulating phases (circles). The coexisting
phases were created after injecting a large tunnel current (5nA).
The inset shows the corresponding $dI/dV-V$ curves.}
\label{figure1}
\end{figure}

To investigate whether we can create controlled phase separation
over a larger area, we carried out LCMAP measurements over a
75nm$\times$75nm area. To start with, we scanned the entire area
at a tunnel current of 1nA and bias of 200mV and recorded the
LCMAP (fig 4(a)). We then increased the tunnel current to 5nA and
scanned the entire area at this higher tunnel current. We then
reduced the tunnel current back to 1nA and recorded the LCMAP
taken over the same area, shown in figure 4(b). We find that, in
this case, the LCMAP is brighter, that is, some regions have
become more conducting upon passing a tunnel current of 5nA. This
implies that by scanning the area at 5nA, the average local
tunneling conductance has increased and this enhanced conductance
state is retained even over a larger area.  We then reversed the
tunnel current and scanned the same region with -5nA and then
recorded the LCMAP with a tunnel current of 1nA again.
Interestingly, this flipping of tunnel current erases the enhanced
conductance and we return to our original starting point as can be
seen in fig 4(c). We find that in general the creation of regions
of higher conductance over a larger area is not homogeneous and
these regions are created generally at steps on the crystal
surface. It is not clear why the preferential destabilization of
the COI state occurs at the steps, however, it may happen that the
charge ordering itself is spatially inhomogeneous and these
regions of the crystal surface make them weaker spots where the
destabilization can be initiated.

\begin{figure}
\begin{center}
\includegraphics[width=8.5cm]{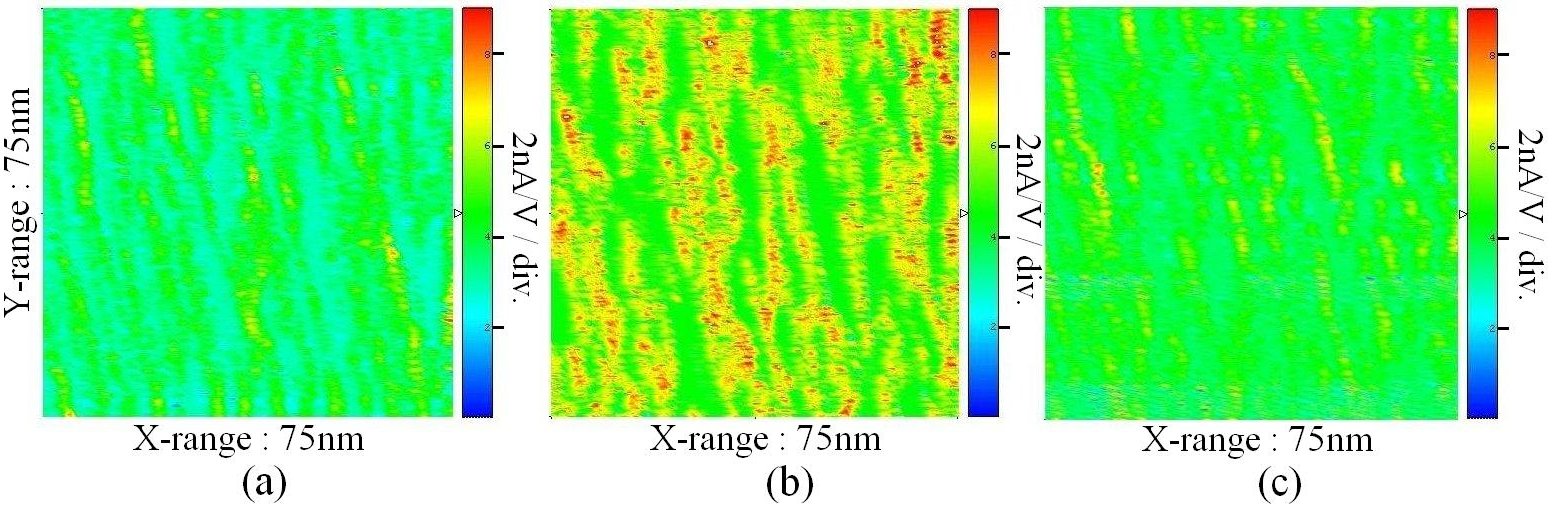}
\end{center}
\vspace{-0.7cm}\caption{Collection of LCMAPs taken at a tunnel
current of 1nA over a 75nm$\times$75nm area, (a) before scanning
the area with 5nA (b) after scanning with 5nA and (c) after having
scanned the area with a tunnel current of -5nA} \label{figure1}
\end{figure}

It should be noted that the underlying topography of the single
crystal surface makes this process somewhat difficult to observe
over larger length scales as most of the contrast in the LCMAP can
sometimes be masked due to the inhomogeneity of the surface. The
experiment in principle can be better done on an epitaxial film.
However, the strain in such films affects the charge ordering and
can initiate phase separation.

     We note that the write-read-erase cycle can be done repeatedly
and also one can reverse the polarity to achieve the same cycle;
that is, one can destabilize the charge ordering with a large
negative tunnel current and subsequently restore it using a tunnel
current of the opposite polarity but having roughly the same
magnitude. This establishes that the direction of the tunnel
current is not important in this process.

     One particular issue that this experiment can address is
whether the phenomenon is driven by current injection or is due to
field effect. We argue that it is current injection driven.
Firstly, since the tunnel junction has much higher resistance
($\approx$ 100M$\Omega$ or more) than the sample, most of the
voltage drop occurs across the tunneling gap and  there is no
significant voltage drop within the sample that can give rise to
the observed phenomenon. Secondly, if the phenomenon is due to a
field effect, it would have depended on the polarity, since this
would decide whether the concentration of holes or electrons was
increasing. It is known that charge ordering is strongest in the
$x=0.5$ composition. Thus, in our sample where $x=0.37$, a
negative bias would imply injection of holes which should drive
the system towards the $x=0.5$ range, making the charge ordering
stronger. Thus destabilization of the COI state would not have
been possible. A positive bias would lead to the opposite effect.
However, we find that regardless of the sign, a higher tunnel
current always destabilizes the charge ordering leading to
formation of the metallic region. In conclusion, we establish that
current injection through an STM tip can give rise to controlled
phase separation thus creating metallic nanoislands in a COI. This
observation raises the possibility of information storage in COI
materials by tunnel current injection.

S.K. thanks CSIR, Government of India, for a fellowship. A.K.R.
thanks DST, Government of India, for a sponsored project.

\pagebreak

\end{document}